\documentclass[letterpaper,australian,american,prl,onecolumn,amsmath,amssymb,superscriptaddress,notitlepage]{revtex4-1}
\usepackage[T1]{fontenc}
\usepackage{xcolor}
\usepackage{babel}
\usepackage{array}
\usepackage{prettyref}
\usepackage{amstext}
\usepackage{graphicx}
\PassOptionsToPackage{normalem}{ulem}
\usepackage{ulem}
\usepackage[unicode=true,
 bookmarks=true,bookmarksnumbered=false,bookmarksopen=false,
 breaklinks=false,pdfborder={0 0 0},pdfborderstyle={},backref=false,colorlinks=false]
 {hyperref}
\hypersetup{pdftitle={Two types of criticality in the brain},
 pdfauthor={Dahmen, Grün, Diesmann, Helias}}
\usepackage{breakurl}

\makeatletter

\providecommand{\tabularnewline}{\\}
\providecolor{lyxadded}{rgb}{0,0,1}
\providecolor{lyxdeleted}{rgb}{1,0,0}

\DeclareRobustCommand{\lyxsout}[1]{\ifx\\#1\else\sout{#1}\fi}

\newrefformat{cap}{\hyperref[#1]{Figure~\ref{#1}}}
\newrefformat{fig}{\hyperref[#1]{Fig.~\ref{#1}}}
\newrefformat{tab}{\hyperref[#1]{Table ~\ref{#1}}}
\newrefformat{sec}{\hyperref[#1]{Section~\ref{#1}}}
\newrefformat{sub}{\hyperref[#1]{Section~\ref{#1}}}
\newrefformat{cha}{\hyperref[#1]{Chapter~\ref{#1}}}
\newrefformat{eq}{\hyperref[#1]{Eq.~(\ref{#1})}}

\definecolor{parametergray}{gray}{0.8}

\usepackage{tabularx} 
\usepackage{multirow}
\usepackage{colortbl}
\usepackage{dsfont}

\makeatother

\begin{document}
\global\long\def\ms{\:\mathrm{ms}}
\global\long\def\mm{\:\mathrm{mm}}
\global\long\def\mum{\:\mathrm{\mu m}}
\global\long\def\mV{\:\mathrm{mV}}
\global\long\def\M{\mathbf{M}}
\global\long\def\I{\mathbf{I}}
\global\long\def\Var{\mathrm{Var}}
\global\long\def\E{\mathrm{E}}
\global\long\def\D{\mathbf{D}}
\global\long\def\a{\mathbf{a}}
\global\long\def\w{\mathbf{w}}
\global\long\def\v{\mathbf{v}}
\global\long\def\erfc{\mathrm{erfc}}
\global\long\def\n{\mathbf{n}}
\global\long\def\q{\mathbf{q}}
\global\long\def\diag{\mathrm{diag}}
\global\long\def\x{x}
\global\long\def\t{\mathbf{t}}
\global\long\def\tr{\mathrm{tr}}
\global\long\def\D{\mathcal{D}}
\global\long\def\bx{x}
\global\long\def\bl{\mathbf{l}}
\global\long\def\bh{\mathbf{h}}
\global\long\def\bJ{J}
\global\long\def\N{\mathcal{N}}
\global\long\def\hh{\hat{h}}
\global\long\def\bhh{\mathbf{\hh}}
\global\long\def\T{\mathrm{T}}
\global\long\def\by{\mathrm{\mathbf{y}}}
\global\long\def\diag{\mathrm{diag}}
\global\long\def\Ftr#1#2{\mathcal{F}\left[#1\right]\left(#2\right)}
\global\long\def\iFtr#1#2{\mathfrak{\mathcal{F}^{-1}}\left[#1\right]\left(#2\right)}
\global\long\def\D{\mathcal{D}}
\global\long\def\T{\mathrm{T}}
\global\long\def\Gammafl{\Gamma_{\mathrm{fl}}}
\global\long\def\gammafl{\gamma_{\mathrm{fl}}}
\global\long\def\E#1{\left\langle #1\right\rangle }
\global\long\def\D{\mathcal{D}}
\global\long\def\bx{\mathbf{x}}
\global\long\def\bl{\mathbf{l}}
\global\long\def\bh{\mathbf{h}}
\global\long\def\bJ{\mathbf{J}}
\global\long\def\N{\mathcal{N}}
\global\long\def\hh{\hat{h}}
\global\long\def\bhh{\mathbf{\hh}}
\global\long\def\T{\mathrm{T}}
\global\long\def\by{\mathrm{\mathbf{y}}}
\global\long\def\D{\mathcal{D}}
\global\long\def\J{\mathbf{J}}
\global\long\def\one{\mathbf{1}}
\global\long\def\e{\mathbf{e}}
\global\long\def\Cpp{\mathcal{K}_{\phi^{\prime}\phi^{\prime}}^{(0)}}
\global\long\def\CCpp{C_{\phi^{\prime}\phi^{\prime}}^{(0)}}
\global\long\def\Cppj{\mathcal{K}_{\phi_{j}^{\prime}\phi_{j}^{\prime}}^{(0)}}
\global\long\def\tx{\tilde{x}}
\global\long\def\xo{x^{(0)}}
\global\long\def\xii{x^{(1)}}
\global\long\def\txi{\tilde{x}^{(1)}}
\global\long\def\bx{x}
\global\long\def\tbx{\tilde{x}}
\global\long\def\bl{j}
\global\long\def\tbj{\tilde{j}}
\global\long\def\bk{\mathbf{k}}
\global\long\def\tbk{\tilde{\mathbf{k}}}
\global\long\def\bh{\mathbf{h}}
\global\long\def\bJ{J}
\global\long\def\bN{\mathcal{N}}
\global\long\def\bH{\mathbf{H}}
\global\long\def\bK{\mathbf{K}}
\global\long\def\bxo{\bx^{(0)}}
\global\long\def\tbxo{\tilde{\bx}^{(0)}}
\global\long\def\bxi{\bx^{(1)}}
\global\long\def\tbxi{\tilde{\bx}^{(1)}}
\global\long\def\tbxi{\tbx^{(1)}}
\global\long\def\tpsi{\tilde{\psi}}
\global\long\def\Cxi{C_{x^{(1)}x^{(1)}}}
\global\long\def\bxxi{\mathbf{\xi}}
\global\long\def\N{\mathcal{N}}
\global\long\def\bW{W}
\global\long\def\bon{\mathbf{1}}
\global\long\def\tj{\tilde{j}}
\global\long\def\tJ{\tilde{J}}
\global\long\def\Z{\mathcal{Z}}
\global\long\def\SOM{S_{\mathrm{OM}}}
\global\long\def\SMSR{S_{\mathrm{MSR}}}
\global\long\def\taum{\tau_{\mathrm{m}}}
\global\long\def\taur{\tau_{\mathrm{r}}}
\global\long\def\EL{E_{\mathrm{leak}}}

\title{Two types of criticality in the brain}

\author{David Dahmen}

\affiliation{Institute of Neuroscience and Medicine (INM-6) and Institute for
Advanced Simulation (IAS-6) and JARA Institute Brain Structure-Function
Relationships (INM-10), Jülich Research Centre, Jülich, Germany}

\author{Sonja Grün}

\affiliation{Institute of Neuroscience and Medicine (INM-6) and Institute for
Advanced Simulation (IAS-6) and JARA Institute Brain Structure-Function
Relationships (INM-10), Jülich Research Centre, Jülich, Germany}

\affiliation{Theoretical Systems Neurobiology, RWTH Aachen University, Aachen,
Germany}

\author{Markus Diesmann}

\affiliation{Institute of Neuroscience and Medicine (INM-6) and Institute for
Advanced Simulation (IAS-6) and JARA Institute Brain Structure-Function
Relationships (INM-10), Jülich Research Centre, Jülich, Germany}

\affiliation{Department of Psychiatry, Psychotherapy and Psychosomatics, Medical
Faculty, RWTH Aachen University, Aachen, Germany}

\affiliation{Department of Physics, Faculty 1, RWTH Aachen University, Aachen,
Germany}

\author{Moritz Helias}

\affiliation{Institute of Neuroscience and Medicine (INM-6) and Institute for
Advanced Simulation (IAS-6) and JARA Institute Brain Structure-Function
Relationships (INM-10), Jülich Research Centre, Jülich, Germany}

\affiliation{Department of Physics, Faculty 1, RWTH Aachen University, Aachen,
Germany}

\date{\today}
\begin{abstract}
Neural networks with equal excitatory and inhibitory feedback show
high computational performance. They operate close to a critical point
characterized by the joint activation of large populations of neurons.
Yet, in macaque motor cortex we observe very different dynamics with
weak fluctuations on the population level. This suggests that motor
cortex operates in a sub-optimal regime. Here we show the opposite:
the large dispersion of correlations across neurons is a signature
of a rich dynamical repertoire, hidden from macroscopic brain signals,
but essential for high performance in such concepts as reservoir computing.
Our findings suggest a refinement of the view on criticality in neural
systems: network topology and heterogeneity endow the brain with two
complementary substrates for critical dynamics of largely different
complexities.
\end{abstract}
\maketitle
The brain is a dynamical system with potentially different regimes
of operation. Network models and experiments suggest optimal computational
performance at critical points, which mark the transition between
two dynamical regimes \citep{Shew13_88}. At critical points, systems
exhibit universal behavior, characterized by strong concerted fluctuations
between all constituents leading to effective long-range interactions
despite short-range couplings. 

A particular type of criticality occurs in neuronal networks with
equal excitatory and inhibitory feedback, leading to neuronal avalanches
\citep{Beggs03_11167} which are visible as large transients of population
activity with a slowly decaying autocorrelation. Signatures of avalanches
have also been observed in mesoscopic measures of neuronal activity
in macaque motor cortex \citep{Petermann09_15921}. However, our parallel
neuronal spiking activity in this cortical region does not show such
large transients of population activity (Fig. 1B,E) and long-range
temporal correlations (Fig. 1F). The data rather show weak and fast
fluctuations of the population activity, suggesting an excess of inhibitory
feedback and dynamically balanced activity \citep{Vreeswijk96}, predicting
low average correlations \citep{Renart10_587,Tetzlaff12_e1002596}
as observed in Fig. 1D.

\begin{figure}
\begin{centering}
\includegraphics{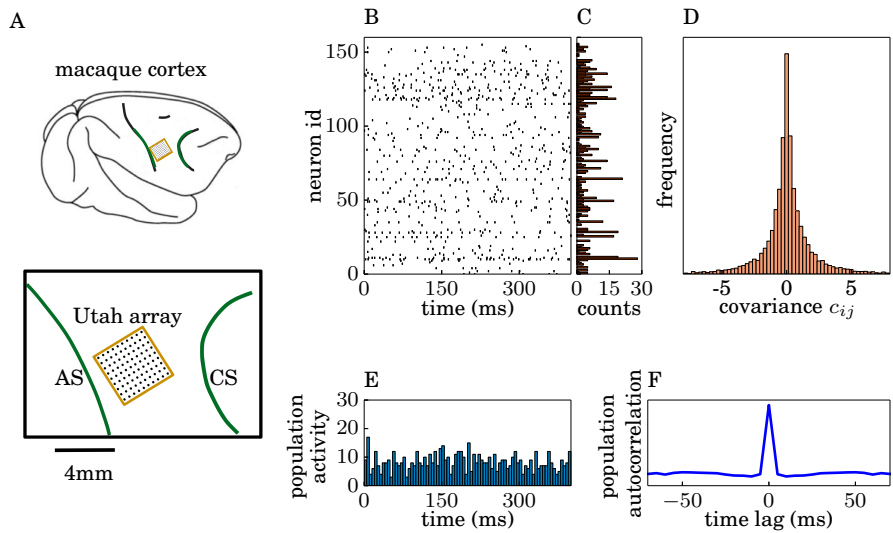}
\par\end{centering}
\caption{\textbf{Spiking activity in macaque motor cortex.} (\textbf{A}) $10\times10$-electrode
``Utah'' array (Blackrock Microsystems, Salt Lake City, UT, USA;
black dots) with $400\,\mathrm{\mu m}$ interelectrode distance covering
an area of $4\times4$ mm$^{2}$ (yellow square) of macaque motor
cortex between the arcuate (AS, left green curve) and the central
sulcus (CS, right green curve) of the right hemisphere. (\textbf{B})
Single trial raster plot of spiking activity of $155$ neurons within
$T=400\,\text{ms}$ after trial start (TS) of a reach-to-grasp task
\citep{Riehle13_48}. (\textbf{C}) Spike counts $n_{i}$ of activity
within $T=400\,\mathrm{ms}$. (\textbf{D}) Distribution of covariances
$c_{ij}=\frac{1}{T}\left(\langle n_{i}n_{j}\rangle-\langle n_{i}\rangle\langle n_{j}\rangle\right)$
between spike counts $n_{i}$ in $141$ trials. (\textbf{E}) Time-resolved
population activity binned in $t_{\mathrm{bin}}=5\,\mathrm{ms}$.
(\textbf{F}) Autocorrelation function of binned population activity.
}
\end{figure}

Another type of criticality, devoid of avalanches, has been investigated
in computational studies \citep{Bertschinger04_1413}: edge-of-chaos
criticality. With increasing heterogeneity in network connections,
the dynamics change from regular to chaotic. At the transition point,
networks possess rich dynamics: a collection of coexisting network
modes leads to topologically complex responses \citep{wainrib13_118101}.
The onset of chaotic dynamics coincides with the breakdown of linear
stability of the deterministic dynamics \citep{Sompolinsky88_259}.
 A direct measurement of the linear stability is, however, hampered
by noise and the sparse sampling of the network dynamics.

\begin{figure}
\begin{centering}
\includegraphics{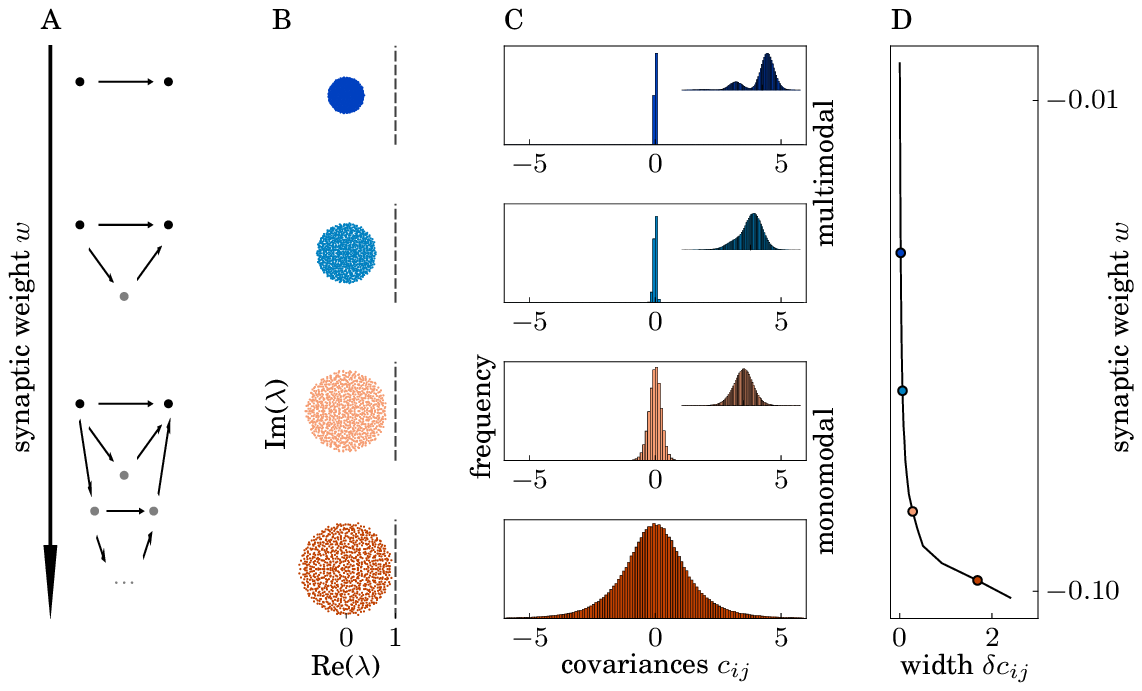}
\par\end{centering}
\caption{\textbf{Dispersion of correlations measures stability of network dynamics.
}Connection strength $w$ in a sparse random network increases from
top to bottom (vertical axis). (\textbf{A}) Connections (arrows) between
a pair of observed neurons (black dots); indirect connections contribute
to correlations via intermediate neurons (gray dots). (\textbf{B})
Bulk eigenvalues $\lambda$ (colored dots) of the connectivity matrix
$W$ in the complex plane (critical line at $\mathrm{Re}(\lambda)=1$).
(\textbf{C}) Distributions of covariances (Eq. 1). Enlargement shown
as insets. (\textbf{D}) Standard deviation $\delta c_{ij}$ of distribution
of covariances (black curve; colored symbols correspond to distributions
shown in panel C). }
\end{figure}

We therefore choose correlations in the neuronal activity (Fig. 1D)
as an indirect measure. They are informative, because they arise from
interactions between neurons, which cause the collective behavior.
Pairwise covariances, to a good approximation, follow the unique and
model-invariant law \citep{Lindner05_061919,Pernice11_e1002059,Trousdale12_e1002408}

\begin{equation}
c(W)=\left[1-W\right]^{-1}D\left[1-W^{\mathrm{T}}\right]^{-1},\label{eq:1}
\end{equation}
which relates the pairwise covariances $c(W)$ to the effective connectivity
matrix $W$ of the linearized dynamics. The latter is the product
of the anatomical connectivity and the sensitivity of individual neurons.
The matrix $D$ can be determined from firing rates, shared and correlated
external inputs, but the final results turn out to be independent
of $D$. 

The linearized network dynamics corresponding to Eq. (1) become unstable
if one eigenvalue of $W$ has a positive real part ($\text{Re}(\lambda)\geq1$).
Eigenvalues are, in principle, determined by all connections in the
network. But their determination from experimentally observed covariances
by Eq. (1) is hindered by severe subsampling; even with massively
parallel recording techniques, at most hundreds of neurons can be
measured simultaneously from the same local network. We therefore
need to find a characteristic of the correlations that is informative
about the eigenvalues, insensitive to the details of the network,
and that can be estimated from a few hundred neurons at most.

We first study theoretically how the distribution of covariances is
affected by the heterogeneity of connections (Fig. 2). In a sparse
network with fixed connection probability, the heterogeneity is uniquely
given by the strength of connections. With increasing connection strength,
activity successively propagates over multiple synapses, leading to
indirectly mediated interactions via a growing number of parallel
paths (Fig. 2A) \citep{Pernice11_e1002059,Trousdale12_e1002408}.
The eigenvalues of the connectivity matrix approach the critical line
where the dynamics lose linear stability (Fig. 2B). Distributions
of covariances (Eq. 1) become monomodal and broader but stay centered
approximately around zero (Fig. 2C,D). These results expose a unique
hallmark of dynamically balanced networks that operate close to linear
instability: widely distributed covariances with a small mean as we
observe in the motor cortex (Fig. 1D). 

\begin{figure}
\begin{centering}
\includegraphics{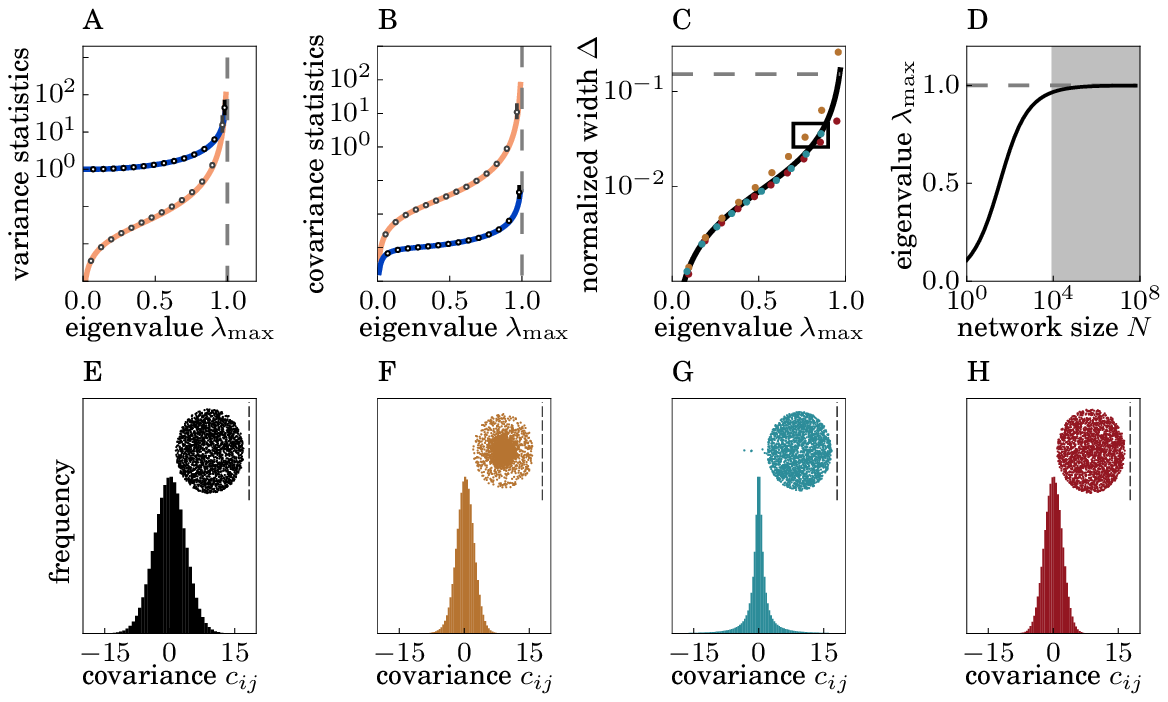}
\par\end{centering}
\caption{\textbf{Large relative dispersion of covariances reveals dynamics
of motor cortex close to instability.} (\textbf{A})-(\textbf{B}) Theoretical
prediction for the mean (Eq. (2), blue curve) and standard deviation
(Eq. (3), orange curve) of variances (panel A) and covariances (panel
B) for different maximum eigenvalues $\lambda_{\mathrm{max}}$ compared
to numerical results (Eq. 1, markers) for homogeneous inhibitory networks.
Ensemble averages for $100$ network realizations (circles), average
upward/downward deviation of the non-Gaussian statistics (bars). Bars
are below marker size except for rightmost data points. (\textbf{C})
Theoretical prediction for the normalized width $\Delta$ (black curve)
of covariances compared to numerical results (Eq. 1, markers) for
excitatory-inhibitory networks (yellow), inhibitory networks with
distance-dependent connectivity (cyan), and numerical simulations
of inhibitory networks of spiking leaky integrate-and-fire neurons
(red). Distributions corresponding to data points within the black
rectangle are shown in panels E-H. (\textbf{D}) Predicted maximum
eigenvalue $\lambda_{\mathrm{max}}$ (Eq. 4) of the effective connectivity
as a function of the number of neurons $N$ for given moments of the
covariance distribution measured in macaque motor cortex ($\Delta=0.15$,
Fig. 1D). The shaded area marks the range of biologically plausible
effective network sizes corresponding to the spatial scale of the
recordings. (\textbf{E})-(\textbf{H}) Distribution of covariances
(histogram) and bulk eigenvalues (dots) relative to the critical line
$\mathrm{Re}(\lambda_{\mathrm{max}})=1$ (dashed black line) for a
homogeneous inhibitory network model (panel E), for a network of excitatory
and inhibitory neurons (panel F), for a network with distance-dependent
connection probability (G), and for a network of spiking leaky integrate-and-fire
neurons (panel H). }
\end{figure}

Although local cortical networks show non-random, cell-type specific,
and distance-dependent connectivity, the simple model of a homogeneous
random network studied in Fig. 2 is sufficient to explain gross features
of the experimental data. To calculate the dispersion of covariances,
we apply a well-established analytical technique for disordered physical
systems: instead of considering all pairwise covariances in a single
network, we observe the covariance of an individual pair of neurons
in different network realizations. The connectivity appears as an
inverse matrix in Eq. (1), which technically complicates the analysis:
no results from random matrix theory apply to this particular problem.
Instead we construct a moment generating function \citep{Chow15}
for the linearized network dynamics \citep{Supplement}, which allows
for the use of spin-glass techniques \citep{DeDominicis78_4913,Sompolinsky82_6860}
combined with approximations for large-$N$ field theories \citep{Moshe03}.
As a result, the disorder contained in the $\sim N^{2}$ entries of
the connectivity formally reduces to only two fluctuating auxiliary
variables that provide input to a fully symmetric all-to-all connected
network. Only this high symmetry and the drastic reduction of dimensionality
enables us to obtain a mean-field theory that describes the neuron-to-neuron
variability \citep{Supplement}. This theory yields the mean and standard
deviation of variances ($i=j$) and covariances ($i\neq j$) to leading-order
in the network size $N$: 
\begin{align}
c_{ij} & =\left[\left[1-\mu\right]^{-1}D_{\lambda}\left[1-\mu^{\mathrm{T}}\right]^{-1}\right]_{ij},\label{eq:1-1}\\
\delta c_{ij} & =\sqrt{\frac{1+\delta_{ij}}{N}\left(\left(\frac{1}{1-\lambda_{\mathrm{max}}^{2}}\right)^{2}-1\right)}D_{\lambda}.\label{eq:1-1-1}
\end{align}
Here, $\mu_{ij}\sim\mathcal{O}(1/\sqrt{N})$ is the mean and $\sigma_{ij}^{2}=\lambda_{\mathrm{max}}^{2}/N\sim\mathcal{O}(1/N)$
is the variance of connection weights in $W$. The latter determines
the radius $\lambda_{\mathrm{max}}$ of the bulk of eigenvalues (Fig.
2B) and the renormalized matrix $D_{\lambda}=D/(1-\lambda_{\mathrm{max}}^{2})$,
which accounts for the structural variability of connections. Eq.
2 predicts that the mean covariances are low ($c_{ij}\sim\mathcal{O}(1/N)$)
if the network is inhibition dominated ($\mu<0$) \citep{Renart10_587,Tetzlaff12_e1002596}.
For large spectral radii $\lambda_{\mathrm{max}}\lesssim1$, Eq. (3),
moreover, predicts a large standard deviation ($\delta c_{ij}\sim\mathcal{O}(1/\sqrt{N})$)
as experimentally observed in our data (Fig. 1D).

The theory, in principle, determines the largest eigenvalue from measured
covariances. But there are two complications: $D_{\lambda}$ is not
known and cannot be measured, and it is unclear how robust the result
is with regard to more realistic connectivity. Both problems are solved
by considering the normalized width of the distribution of covariances
$\Delta=\delta c_{ij}/c_{ii}$. This measure is predominantly determined
by the network size and the most unstable eigenvalue, $\lambda_{\mathrm{max}}$
(Fig. 3C). The prediction of $\Delta$ is sufficient even for network
topologies such as excitatory-inhibitory networks (Fig. 3F) and distance-dependent
connection probabilities (Fig. 3G). The latter networks also qualitatively
explain the shape of the experimentally observed covariance distribution.
The applicability of the theory goes beyond linear network dynamics;
it predicts $\Delta$ even in spiking networks (Fig. 3H). Therefore,
the normalized width $\Delta$ can be used to infer the operational
regime of the cortical network (Fig. 1). The distance to linear instability
is determined by $\lambda_{\mathrm{max}}$ which, to leading order
in $N$, is given by inversion of Eqs. (2) and (3) as 
\begin{equation}
\lambda_{\mathrm{max}}=\sqrt{1-\sqrt{\frac{1}{1+N\Delta^{2}}}}.\label{eq:spectral_radius}
\end{equation}
The dispersion measured in massively parallel spike recordings of
macaque motor cortex (Fig. 1D) predicts that the network operates
close to instability (Fig. 3D). Biologically plausible neuron numbers
$N$ below the recording array are above $10^{4}$. Together with
the measured relative width $\Delta=0.15$ (with bias correction due
to the finite amount of measured data \citep{Supplement}) this leads
to a small quantity $N\Delta^{2}$ , such that Eq. \eqref{eq:spectral_radius}
predicts $\lambda_{\mathrm{max}}\lesssim1$ (Fig. 3D, gray area).

In general, a value $\lambda_{\mathrm{max}}\lesssim1$ results from
the large heterogeneity of connections across neuron pairs. It implies
a large number of eigenvalues of the effective connectivity matrix
being close to the critical line where linear stability breaks down
(Fig. 4A). Due to the proximity of the eigenvalues to this critical
line, the network possesses a rich dynamical repertoire of multiple-neuron
responses (modes) with different time courses (Fig. 4C). The contribution
of each neuron to a particular mode is different for each neuron and
determined by the eigenvector of the connectivity matrix corresponding
the mode (Fig. 4A, cyan bars). Many modes have large time constants
that would be visible in slowly decaying autocorrelation functions
(Fig. 4C). However, a direct experimental identification of these
modes is a major challenge. The often-considered population activity
is only one particular mode where all neurons contribute equally.
The almost vanishing mean of the covariances (Fig. 1D, Fig. 4B) and
the weakly fluctuating and quickly decaying population activity (Fig.
1F, Fig. 4C) show that the corresponding population eigenvalue (Fig.
4A, yellow dot) is negative, the feedback is inhibition dominated
and the network is said to be dynamically balanced (\citep{Renart10_587,Cohen11_811,Tetzlaff12_e1002596},
Fig. 4, Case 1). 

This operational regime is in contrast to avalanche criticality \citep{Tkacik15_11508}
in networks with equal excitatory and inhibitory feedback ($\mu\approx1/N$,
Fig. 4, Case 2). The strongly fluctuating population activity observed
in such networks causes positive covariances (Fig 4E) and a slowly
decaying autocorrelation function of the population activity (Fig.
4F). These dynamics are determined by the single, nearly unstable
eigenvalue of the population activity that results from the average
connectivity structure of the network \citep{Larremore11_058101}
(Fig 4D, yellow dot). Each of the remaining $N-1$ modes has a low
amplitude and an exponentially decaying, fast dynamics (Fig. 4F).
The network in such a critical state is hence effectively one-dimensional
\citep{Tkacik14_e1003408}.

The high-dimensional criticality found here is not necessarily specific
to macaque motor cortex. Experimental evidence for the operation of
cortical networks in the dynamically balanced state is overwhelming
\citep{Okun2008_535,Reinhold15_1789,Dehghani16}. In addition to the
low average covariances in this state, other cortical areas, such
as visual cortex \citep{Ecker10}, also show a covariance dispersion
comparable to our data \citep{Supplement}. The same analysis applied
to different cortical areas and experimental conditions can be used
to determine their operational point $\lambda_{\mathrm{max}}$.

\begin{figure}
\begin{centering}
\includegraphics{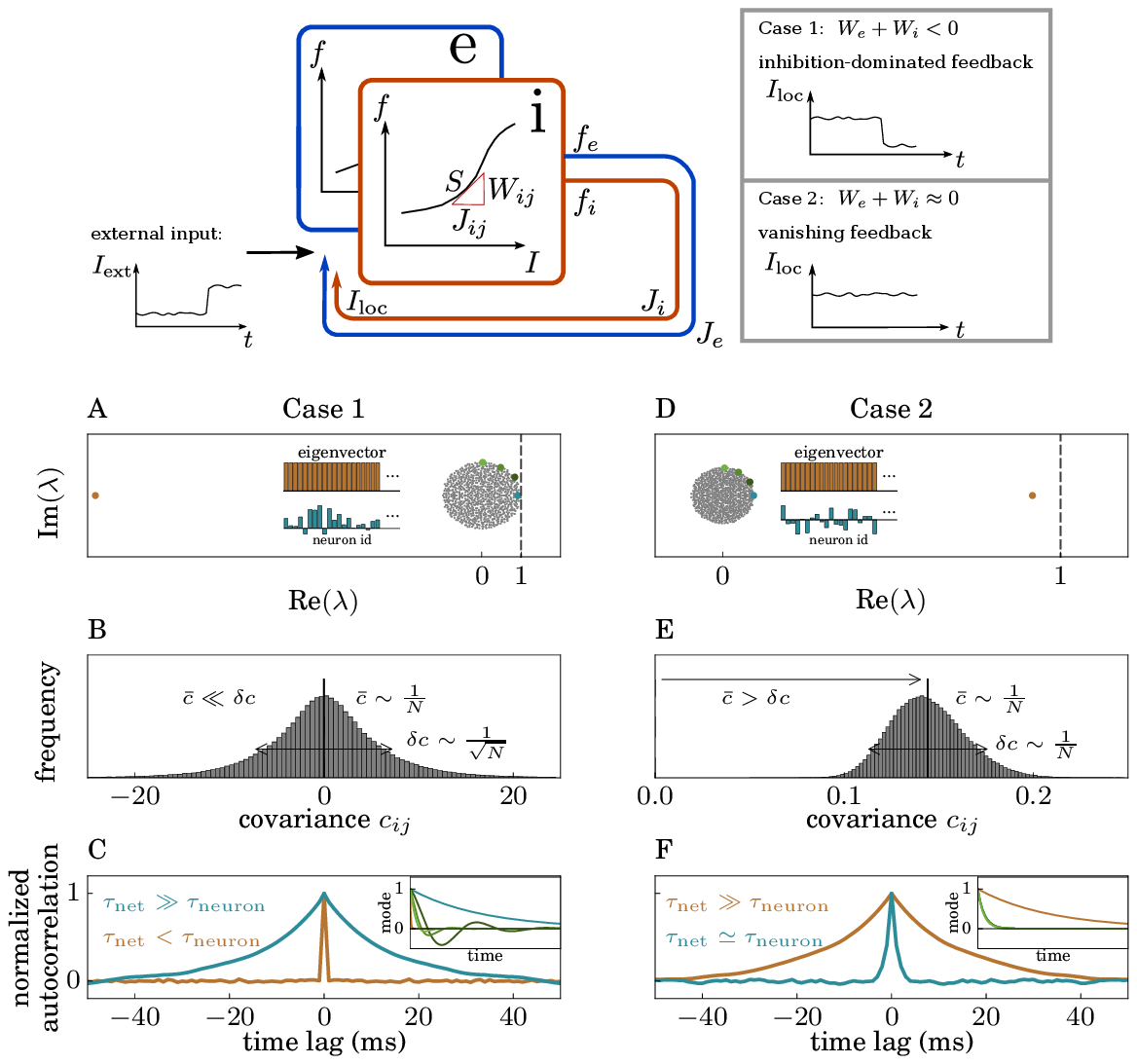}
\par\end{centering}
\caption{\textbf{Two types of criticality.} \textbf{Top}: Classification of
networks based on feedback of population activity. A perturbation
by an external input $\delta I_{\mathrm{ext}}$ (black time course)
to an excitatory (blue) and inhibitory (red) population of neurons
causes changes in firing rates $\delta f=S\cdot\delta I_{\mathrm{ext}}$
and recurrent inputs $\delta I_{\mathrm{loc}}=J_{e}\delta f_{e}+J_{i}\delta f_{i}=(W_{e}+W_{i})\delta I_{\mathrm{ext}}$
via excitatory ($J_{e}$, blue arrow) and inhibitory ($J_{i},$ red
arrow) local connections. The recurrent input either counteracts the
change in the external input (Case 1, inhibition-dominated feedback,
top gray box), remains unaffected (Case 2, vanishing feedback, bottom
gray box) or amplifies the external perturbation (locally unstable,
not shown). \textbf{Bottom}: Resulting activity statistics on the
level of individual neurons in cases 1 and 2, respectively, in the
critical regime. Case 1 (left): Dynamically balanced network with
stable population activity but virtually unstable linearized dynamics
hidden in specific linear combinations of neuron activities. (\textbf{A})
Spectrum of eigenvalues with negative outlier (yellow dot) and nearly
critical bulk eigenvalue (cyan dot) and corresponding eigenvector
(cyan bars) generated by heterogeneity in connections across neuron
pairs. (\textbf{B}) Distribution of covariances with almost vanishing
mean and large standard deviation. (\textbf{C}) Autocorrelation functions
of the population activity (yellow curve, cf. yellow dot in panel
A) and the activity projected onto the eigenvector corresponding to
the largest real bulk eigenvalue (cyan curve). Inset: Time course
of network modes corresponding to colored eigenvalues in panel A.
Case 2 (right): Network with almost vanishing excitatory and inhibitory
feedback and virtually unstable linearized population dynamics. (\textbf{D})
Spectrum of eigenvalues with positive outlier (yellow dot) and corresponding
eigenvector (yellow bars, population activity) generated by average
connectivity structure. (\textbf{E}) Distribution of covariances with
positive mean and small standard deviation. (\textbf{F}) Autocorrelation
functions of the population activity (yellow curve) and the activity
projected onto the eigenvector corresponding to the largest bulk eigenvalue
(cyan curve, cf. cyan dot in panel D). Inset: Time course of network
modes corresponding to colored eigenvalues in panel D. }
\end{figure}

The operation in the dynamically balanced, critical regime ($\lambda_{\mathrm{max}}\lesssim1$)
suggests several implications for learning and information processing.
Neurons belonging to a critical mode show pairwise covariances that
strongly exceed the average (Fig. 4B). Such covariances are likely
to interact with spike-timing dependent synaptic plasticity \citep{Markram97a,Morrison08_459},
leading to an increased interaction between neuronal and synaptic
dynamics. Weak external inputs to the network are, moreover, sufficient
to shift a large number of eigenvalues across the edge of stability
\citep{wainrib13_118101}, and thereby drastically change the recurrent
network dynamics. Critical modes have a multitude of characteristic
shapes and life times (Fig. 4C inset, \citep{Hennequin14_1394});
they arise here despite the stereotypical and fast dynamics of individual
neurons as a result of the heterogeneity of the network. The rich
repertoire enables the parallel integration and maintenance of signals
over prolonged time scales. Such networks provide a wealth of transformations
on the input and therefore may serve as an exhaustive reservoir for
computation \citep{Maass02_2531,Jaeger04_87}. 

Finally, the two types of criticality are not mutually exclusive as
they are governed by different mechanisms. They can coexist in different
brain regions or even in the same local network; networks may hence
dynamically be moved into either regime to adapt brain function to
momentary demands.

\section*{Acknowledgments }

We are grateful to Alexa Riehle and Thomas Brochier for providing
the experimental data. We thank Vahid Rostami for help with the analysis
of the multi-channel data. The research was carried out within the
scope of the International Associated Laboratory ``Vision for Action
- LIA V4A`` of INT (CNRS, AMU), Marseilles and INM-6, Jülich. This
work was partially supported by HGF young investigator's group VH-NG-1028,
Helmholtz portfolio theme SMHB, and EU Grants 604102 and 720270 (Human
Brain Project, HBP).

\newpage{}

\part*{Supplementary Materials}

\setcounter{figure}{0} 
\renewcommand{\thefigure}{S\arabic{figure}}
\renewcommand{\thetable}{T\arabic{table}}

\section{Mean-field theory for meta-statistics beyond self-averaging\label{sec:Mean-field-theory-for}}

Mean-field theory most-commonly employs the thermodynamic limit ($N\to\infty$),
reducing the collective dynamics of the $N$ interacting units to
$N$ pairwise independent units, each subject to a self-consistently
determined auxiliary field \citep{Sompolinsky88_259,Vreeswijk96,Aljadeff15_088101,Kadmon15_041030}.
Covariances of individual neurons are self-averaging in this limit,
so they are identical for all units and independent of the realization
of the randomness in the network connectivity. In particular, cross-covariances
vanish. In the absence of external stimuli and in the weaky correlated
regime, covariances can be understood in linear response theory \citep{Grytskyy13_131}.
For such a linearized network model, interactions between neurons
can be included as a finite-size correction within this self-averaging
framework \citep{Ginzburg94,Renart10_587,Trousdale12_e1002408,Tetzlaff12_e1002596,Helias14}.
While this procedure yields covariances averaged over many pairs of
units, the experimentally observed neuron to neuron variability \citep{Ecker10}
is lost. Here, we develop a theory beyond self-averaging covariances
that captures their statistics. It exploits the large size of biologically
realistic local networks, but includes finite-size fluctuations of
auxiliary fields which derive from the quenched disorder of the couplings
to explain the variability of covariances. To perform this qualitative
step, we need to combine methods from different fields: We use a functional
formulation of Gaussian processes that originates from statistical
field theory \citep{DeDominicis78_4913,Chow15,Hertz16_033001} and
combine it with methods typically used for disordered systems in the
large $N$ limit \citep{Moshe03}, such as spin glasses \citep{Sompolinsky82_6860,Fischer91c}.
We outline these steps in detail below.

\subsection{Moment generating functional for the network dynamics}

In the following, we consider time-lag integrated covariances 
\begin{equation}
c_{ij}=\int_{-\infty}^{\infty}c_{ij}(\tau)d\tau=\int_{-\infty}^{\infty}\int_{-\infty}^{\infty}\left\langle x_{i}(t+\tau)x_{j}(t)\right\rangle _{x}dt\,d\tau=\left\langle X_{i}(0)X_{j}(0)\right\rangle _{x},\label{eq:corr_stats_int_cov}
\end{equation}
calculated as averages $\left\langle \right\rangle _{x}$ across different
trials of the linearized network dynamics $x(t)$ or its Fourier transform
$X(\omega)$ evaluated at $\omega=0$ (Wiener-Khinchin theorem, \citep[sec. 1.4.2]{Gardiner85}).
The linearized dynamics can be modeled as a set of coupled Ornstein-Uhlenbeck
processes

\begin{equation}
\tau\,\frac{dx(t)}{dt}=-x(t)+W\cdot x(t)+\xi(t)\label{eq:OUP_functional}
\end{equation}
with the moment-generating functional
\begin{align}
Z[\bl] & =\int\mathcal{D}\bx\int\mathcal{D}\tbx\,\exp\Big(S_{0}[\bx,\tbx]+\bl^{\T}\bx\Big)\nonumber \\
\text{with }\quad S_{0}[\bx,\tbx] & =\tbx^{\T}\left(\partial_{t}+1-W\right)\,\bx+\frac{D}{2}\tbx^{\T}\tbx.
\end{align}
Here $\xi(t)$ is a Gaussian white noise with variance $\left\langle \xi_{i}(t)\xi_{j}(t^{\prime})\right\rangle =D\delta_{ij}\delta(t-t^{\prime})$,
$\tbx(t)$ is a purely imaginary response field, $\int\mathcal{D}\tilde{x},\,\int\mathcal{D}x$
are suitably defined path integral measures, and $\tbx^{\T}\tbx=\sum_{i}\int\,\tilde{x}_{i}(t)\,\tilde{x}_{i}(t)\,dt$
is a scalar product \citep{DeDominicis78_4913,Chow15,Hertz16_033001}.
The generating functional can easily be interpreted in Fourier domain
due to the linearity of \prettyref{eq:OUP_functional} and the invariance
of scalar products under unitary transforms
\begin{align}
Z[J] & =\int\mathcal{D}X\int\mathcal{D}\tilde{X}\,\exp\Big(S_{0}[X,\tilde{X}]+J^{\T}X\Big)\nonumber \\
\text{with }S_{0}[X,\tilde{X}] & =\tilde{X}^{\T}\left(i\omega+1-W\right)\,X+\frac{D}{2}\tilde{X}^{\T}\tilde{X},
\end{align}
with Fourier transformed variables denoted by capital letters. The
scalar product in frequency domain reads $\tilde{X}^{\T}X=\sum_{i}\int\,\tilde{X}_{i}(-\omega)X_{i}(\omega)\,d\omega$.
The generating functional factorizes into generating functions for
each frequency $\omega$. As shown in \prettyref{eq:corr_stats_int_cov},
time-lag integrated covariances only require the knowledge of $X(0)$.
In the following, we will therefore only discuss zero frequency. After
integration over all non-zero frequencies one obtains the generating
function for zero frequency 
\begin{align}
Z(J) & =\det(1-W)\int\mathcal{D}X\int\mathcal{D}\tilde{X}\,\exp\Big(S_{0}(X,\tilde{X})+J^{\T}X\Big)\label{eq:ZJ}\\
 & =\exp\left(\frac{1}{2}J^{\mathrm{T}}\left(1-W\right)^{-1}D\left(1-W^{\mathrm{T}}\right)^{-1}J\right)\nonumber \\
\text{with }\quad S_{0}(X,\tilde{X}) & =\tilde{X}^{\T}\left(1-W\right)\,X+\frac{D}{2}\tilde{X}^{\T}\tilde{X},\label{eq:action-1}
\end{align}
with the single-frequency ($\omega=0$) scalar product defined as
$\tilde{X}^{\T}X=\sum_{i}\tilde{X}_{i}X_{i}$, and integration measures
$\int\mathcal{D}X=\prod_{j}\int_{-\infty}^{\infty}dX_{j}$ and $\int\mathcal{D}\tilde{X}=\prod_{j}\frac{1}{2\pi i}\int_{-i\infty}^{i\infty}d\tilde{X}_{j}$.
The determinant in \prettyref{eq:ZJ} follows from the normalization
condition $Z[J=0]=1$. The time-lag integral of the covariance functions
follows as 

\begin{alignat}{1}
c(W) & =\left[1-W\right]^{-1}D\left[1-W^{\mathrm{T}}\right]^{-1}.\label{eq:covs_corr_stats}
\end{alignat}

\subsection{Self-averaging meta-statistics}

Equation \eqref{eq:covs_corr_stats} relates covariances between individual
pairs of neurons to the connectivity matrix $W$. These, however,
change between realizations of the random connectivity. In contrast,
the meta-statistics, by which we denote the moments of the distribution
of covariances, can be assumed constant across realizations (self-averaging,
\citep{Fischer91c}). We therefore seek for an expression relating
the moments of the distribution of covariances to the statistics of
the connectivity $W$. 

We denote with $\bar{}$ the empirical average, with $\langle\rangle_{x}$
the expectation over realizations of the processes $x$, and with
$\langle\rangle$ the ensemble average over the disordered connectivity
$W$. Exchanging the order of differentiation and averaging allows
expressing second moments of covariances as derivatives of a single
disorder-averaged generating function $\left\langle Z(J)\right\rangle $:
We first assume the empirical average to be self-averaging
\begin{equation}
\overline{c_{ii}^{2}}=\left\langle \overline{c_{ii}^{2}}\right\rangle .
\end{equation}
We then use its definition as $\overline{c_{ii}^{2}}=\frac{1}{N}\sum_{i=1}^{N}c_{ii}^{2}$,
and that of the second cumulant $c_{ij}=\left\langle X_{i}X_{j}\right\rangle _{x}$
of the zero frequency components $X=X(\omega=0)=\int_{-\infty}^{\infty}x(t)\,dt$
of Ornstein-Uhlenbeck processes $x$ to obtain
\begin{equation}
\overline{c_{ii}^{2}}=\frac{1}{N}\sum_{i=1}^{N}\left\langle \left\langle X_{i}X_{i}\right\rangle _{x}^{2}\right\rangle .\label{eq:squared_second}
\end{equation}
As the action \prettyref{eq:ZJ} for a single realization of $W$
is quadratic, Wick's theorem applies, $\left\langle X_{i}X_{i}X_{j}X_{j}\right\rangle _{x}=2\left\langle X_{i}X_{j}\right\rangle _{x}^{2}+\left\langle X_{i}X_{i}\right\rangle _{x}\left\langle X_{j}X_{j}\right\rangle _{x}=2c_{ij}^{2}+c_{ii}c_{jj}$
for the special case $i=j$, such that we may identify the squared
second moment \eqref{eq:squared_second} with a fourth moment. The
latter can be expressed with the help of the generating function
\begin{eqnarray}
\overline{c_{ii}^{2}} & = & \frac{1}{N}\sum_{i=1}^{N}\left\langle \left\langle X_{i}X_{i}\right\rangle _{x}^{2}\right\rangle \stackrel{\text{Wick's th.}}{=}\frac{1}{N}\sum_{i=1}^{N}\frac{1}{3}\left\langle \left\langle X_{i}X_{i}X_{i}X_{i}\right\rangle _{x}\right\rangle \nonumber \\
 & = & \frac{1}{N}\sum_{i=1}^{N}\frac{1}{3}\left.\left\langle \frac{d^{4}}{dJ_{i}^{4}}Z(J)\right\rangle \right|_{J=0}=\frac{1}{3}\left.\frac{d^{4}}{dJ_{i}^{4}}\left\langle Z(J)\right\rangle \right|_{J=0}.\label{eq:c_ii2_bar-1}
\end{eqnarray}
In the last step, we exchanged the order of derivatives and the expectation
value over network realizations and used the symmetry of the disorder-averaged
network over units. Analogously follows for $i\neq j$ and $N-1\approx N$,
\begin{equation}
\overline{c_{ij}^{2}}=\frac{1}{2}\left.\frac{d^{4}}{dJ_{i}^{2}dJ_{j}^{2}}\left\langle Z(J)\right\rangle \right|_{J=0}-\frac{1}{2}\langle\overline{c_{ii}}\rangle^{2}.\label{eq:c_ij2_bar-1}
\end{equation}

\subsection{Disorder-averaged generating function}

Ignoring insignificant variations in the normalization $\det(1-W)$
of $Z[J]$, the disorder average only affects the coupling term in
\prettyref{eq:ZJ}
\begin{eqnarray*}
\left\langle \exp\left(\tilde{X}^{\mathrm{T}}W\,X\right)\right\rangle  & = & \left\langle \exp\left(\sum_{i,j}W_{ij}\tilde{X}_{i}X_{j}\right)\right\rangle \\
 & = & \prod_{i,j}\left\langle \exp\left(W_{ij}\tilde{X}_{i}X_{j}\right)\right\rangle \\
 & = & \prod_{i,j}\exp\left(\sum_{k=1}^{\infty}\frac{\kappa_{k}}{k!}(\tilde{X}_{i}X_{j})^{k}\right).
\end{eqnarray*}
For clarity, we here assume independent and identically distributed
weights $W_{ij}$. In the resulting cumulant expansion \citep{Sherrington75_1792,DeDomincis78_353,Nishimori01_01,Fischer91c}
$\kappa_{k}$ is the $k$-th cumulant for a single connection $W_{ij}$
\citep{Gardiner83}. For fixed connection probability $p$, the number
of inputs to a neuron scales with the network size $N$. To keep the
input and its fluctuations within a certain dynamic range when increasing
the network size, we require synaptic weights to scale with $1/\sqrt{N}$
\citep{Vreeswijk96,Vreeswijk98}, such that the cumulant expansion
is an expansion in $1/\sqrt{N}$. A truncation at the second cumulant
($\propto N^{-1}$) maps $W$ to a Gaussian connectivity $\mathcal{N}(\mu,\lambda_{\mathrm{max}}^{2}/N)$
so that 
\begin{eqnarray}
\left\langle Z(J)\right\rangle  & \sim & \int\mathcal{D}X\int\mathcal{D}\tilde{X}\,\exp\Big(S_{0}(X,\tilde{X})+\frac{\lambda_{\mathrm{max}}^{2}}{2N}V(X,\tilde{X})+J^{\T}X\Big),\label{eq:corr_stats_Sint}\\
S_{0}(X,\tilde{X}) & = & \tilde{X}^{\T}\left(1-\mu\right)X+\frac{D}{2}\tilde{X}^{\T}\tilde{X},\nonumber \\
V(X,\tilde{X}) & = & \tilde{X}^{\T}\tilde{X}\,X^{\mathrm{T}}X,\nonumber 
\end{eqnarray}
with a homogeneous mean connection weight $\mu_{ij}=\mu=\mathcal{O}(1/\sqrt{N})$.
The second cumulant ($\lambda_{\mathrm{max}}^{2}/N$) is the first
non-trivial contribution to the second moment of covariances. While
higher cumulants of the connectivity have an impact on higher moments
of the distribution of covariances, their effect on the first two
moments is suppressed by the large network size.

\subsection{Auxiliary-field formalism}

\begin{figure}
\centering{}\includegraphics[width=0.8\textwidth]{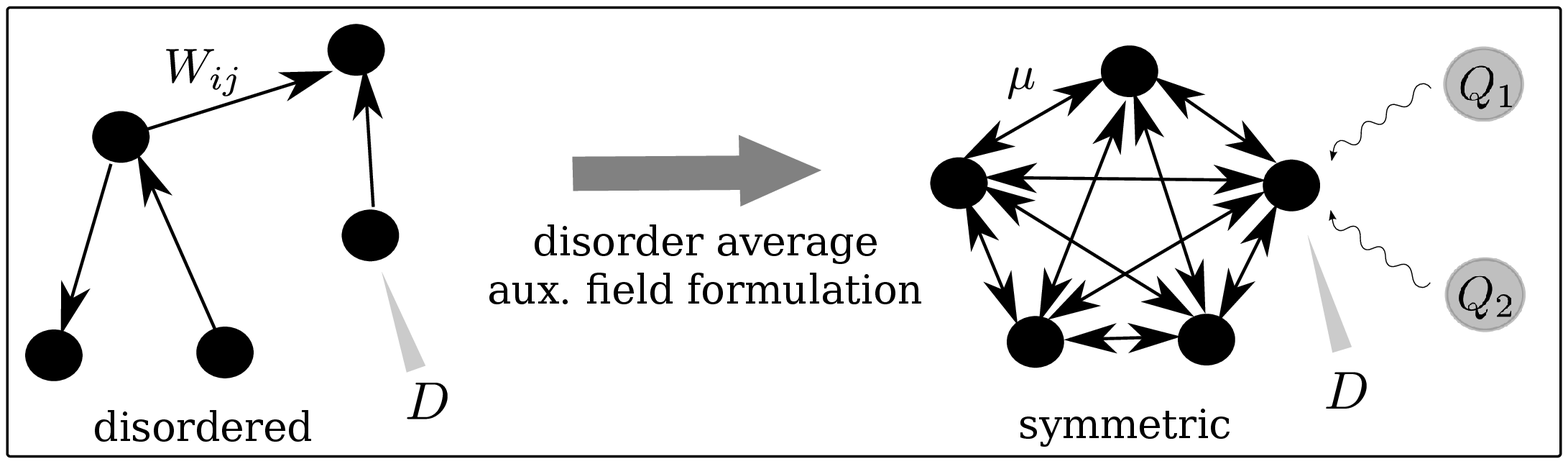}\caption{\textbf{Reduction of a disordered to a regular network.} Disorder
average maps network with frozen variability in connections (left)
to highly symmetric network on the background of fluctuating auxiliary
fields $Q$ (right). Their fluctuations contribute globally to the
covariance that drives the network fluctuations (illustrated by wavy
arrows).\label{fig:mapping} }
\end{figure}

The interaction term $V$ prevents an exact calculation of the disorder-averaged
generating function. A converging perturbation series can be obtained
in the auxiliary-field formulation \citep{Moshe03}, where a field
$Q_{1}=\frac{\lambda_{\mathrm{max}}^{2}}{N}X^{\mathrm{T}}X$ is introduced
for the sum of a large number of statistically equivalent activity
variables. Using the Hubbard-Stratonovich transformation 
\begin{eqnarray*}
e^{\frac{\lambda_{\mathrm{max}}^{2}}{2N}\tilde{X}^{\T}\tilde{X}\,X^{\mathrm{T}}X} & = & \int_{-\infty}^{\infty}dQ_{1}\delta\left(Q_{1}-\frac{\lambda_{\mathrm{max}}^{2}}{N}X^{\mathrm{T}}X\right)\,e^{\frac{1}{2}Q_{1}\tilde{X}^{\T}\tilde{X}}\\
 & = & \underbrace{\frac{1}{2\pi i}\frac{N}{\lambda_{\mathrm{max}}^{2}}\int_{-\infty}^{\infty}dQ_{1}\int_{-i\infty}^{i\infty}dQ_{2}}_{\int\mathcal{D}Q}e^{-\frac{N}{\lambda_{\mathrm{max}}^{2}}Q_{1}Q_{2}+\frac{1}{2}Q_{1}\tilde{X}^{\T}\tilde{X}+Q_{2}X^{\mathrm{T}}X},
\end{eqnarray*}
one obtains a free theory, which is a quadratic action in the activity
($X$) and response variables ($\tilde{X})$, on the background of
fluctuating fields $Q$
\begin{eqnarray}
\left\langle Z(J)\right\rangle  & \sim & \int\mathcal{D}Q\,\exp\left(-\frac{N}{\lambda_{\mathrm{max}}^{2}}Q_{1}Q_{2}+\ln\left(Z_{Q}(J)\right)\right),\label{eq:mean_Z}\\
Z_{Q}(J) & = & \int\mathcal{D}X\int\mathcal{D}\tilde{X}\;\exp\left(S_{0}(X,\tilde{X})+\frac{1}{2}Q_{1}\tilde{X}^{\T}\tilde{X}+Q_{2}X^{\mathrm{T}}X+J^{\T}X\right).\nonumber 
\end{eqnarray}
The high dimensional integrals of the free theory $Z_{Q}(J)$ can
be solved analytically yielding a two-dimensional interacting theory
in the auxiliary fields $Q_{1}$ and $Q_{2}$. The auxiliary field
formalism translates the high-dimensional ensemble average over $W$
to a low-dimensional average over $Q$; it maps the local disorder
in the connections to fluctuations of global fields $Q$ interacting
with a highly symmetric all-to-all connected network, illustrated
in \prettyref{fig:mapping}. Only in the special case of vanishing
mean connection strength $\mu=0$, the system factorizes into $N$
unconnected units, each interacting with the same set of fields $Q$.
The all-to-all network not only captures the autocovariance of a single
neuron, but also the cross-covariance with any other neuron.

\subsection{Saddle-point approximation}

Performing a change of variables $Q=Q^{*}(J)+\frac{\delta Q}{\sqrt{N}}$
in \prettyref{eq:mean_Z}, where $Q^{*}(J)$ are the saddle-points
of the exponent $-\frac{N}{\lambda_{\mathrm{max}}^{2}}Q_{1}Q_{2}+\ln\left(Z_{Q}(J)\right)=:-N\cdot Y(Q_{1},Q_{2},J)$
determined by the saddle-point equations
\begin{eqnarray}
0 & = & \left.\frac{\partial}{\partial Q_{\alpha}}Y(Q_{1},Q_{2},J)\right|_{Q=Q^{*}(J)}\quad\alpha\in\{1,2\},\label{eq:saddle_1-1-1}
\end{eqnarray}
and expanding $Y(Q_{1},Q_{2},J)$ around $Q^{*}(J)$, yields an expansion
in powers of $1/N$ \citep{Moshe03,Amit05}. We focus only on the
leading-order contribution $Y(Q_{1}^{*}(J),Q_{2}^{*}(J),J)$ which
describes tree-level diagrams in the $Q$-theory and drop all source
dependence in higher Taylor coefficients 
\begin{equation}
\left\langle Z(J)\right\rangle \sim\exp\left(-N\cdot Y(Q_{1}^{*}(J),Q_{2}^{*}(J),J)\right),\label{eq:Z-Y}
\end{equation}
with the corresponding action
\begin{equation}
S_{Q_{\alpha}^{\ast}(J)}(X,\tilde{X})=\tilde{X}^{\mathrm{T}}\,\left(1-\mu\right)\,X+\frac{D+Q_{1}^{*}(J)}{2}\,\tilde{X}^{\T}\tilde{X}+Q_{2}^{*}(J)\,X^{\mathrm{T}}X.\label{eq:action_saddle_point}
\end{equation}
Note that this Gaussian theory in $X,\tilde{X}$ still contains in
the latter two terms contributions from the quartic interaction term
$V$ of the original theory. Variability in $Q^{*}(J)$, through their
source dependence, therefore gives rise to cumulants of the activity
variables beyond second order. The dependence of $Q^{*}(J)$ on external
sources $J$ was neglected in prior work \citep{Sompolinsky82_6860}
since it scales to leading order as $1/N$. This approximation, however,
yields a Gaussian theory (see \prettyref{eq:action_saddle_point}
for $J=0$), which does not generate fourth cumulants and therefore
no distribution of covariances. Taking into account the $J$ dependence
of auxiliary fields in combination with the relation between fourth
cumulants and distributions of covariances \prettyref{eq:c_ii2_bar-1}
and \prettyref{eq:c_ij2_bar-1} thus extends mean-field theory beyond
self-averaging covariances. 

The action \eqref{eq:action_saddle_point} shows that $Q_{1}^{*}(J)$
acts as a global contribution to the Gaussian noise whereas $Q_{2}^{*}(J)$
directly contributes to the inverse of the covariance matrix. By integrating
out the response variables $\tilde{X}$, one can alternatively interpret
both $Q_{\alpha}{}^{*}(J)$ as contributions to the covariance matrix
of the noise. 

Saddle points $Q^{*}(J)$ are obtained self-consistently from \prettyref{eq:saddle_1-1-1}
which reduces to the set of equations
\begin{eqnarray}
Q_{1}^{*}(J) & = & \frac{\lambda_{\mathrm{max}}^{2}}{N}\,\left\langle X^{\mathrm{T}}X\right\rangle _{Q_{\alpha}^{\ast}(J)}(J)\nonumber \\
Q_{2}^{*}(J) & = & \frac{\lambda_{\mathrm{max}}^{2}}{2N}\,\left\langle \tilde{X}^{\T}\tilde{X}\right\rangle _{Q_{\alpha}^{\ast}(J)}(J)\nonumber \\
\text{with}\qquad\left\langle \circ\right\rangle _{Q_{\alpha}^{\ast}(J)}(J) & := & \frac{\int\mathcal{D}X\int\mathcal{D}\tilde{X}\:\circ\,\exp\left(S_{Q_{\alpha}^{\ast}(J)}(X,\tilde{X})+J^{\T}X\right)}{\int\mathcal{D}X\int\mathcal{D}\tilde{X}\:\exp\left(S_{Q_{\alpha}^{\ast}(J)}(X,\tilde{X})+J^{\T}X\right)}\label{eq:self-consis}
\end{eqnarray}
The above set of equations cannot be solved analytically for $Q^{*}(J)$.
However, moments of activity result from derivatives of $\left\langle Z(J)\right\rangle $
evaluated at $J=0$. The derivatives act not only on the source term
$J^{\mathrm{T}}X$, but also on the $J$-dependence of the saddle-point.
Therefore, moments are determined by saddle-points and their derivatives
evaluated at zero source. Setting $J=0$ in \prettyref{eq:self-consis}
yields the self-consistent result
\begin{eqnarray}
\left\langle \tilde{X}_{i}\tilde{X}_{j}\right\rangle _{Q_{\alpha}^{\ast}} & = & 0,\\
\left\langle X_{i}X_{j}\right\rangle _{Q_{\alpha}^{\ast}} & = & \left[\left(1-\mu\right)^{-1}\left(D+Q_{1}^{*}\right)\left(1-\mu^{\mathrm{T}}\right)^{-1}\right]_{ij},\label{eq:2ptcorreletor-1}\\
\left\langle \tilde{X}_{i}X_{j}\right\rangle _{Q_{\alpha}^{\ast}} & = & \left[\left(1-\mu\right)^{-1}\right]_{ij},
\end{eqnarray}
with 
\begin{align}
Q_{1}^{*}:=Q_{1}^{*}(0) & =\frac{R^{2}}{1-R^{2}}\,D,\label{eq:corr_stats_Q0}\\
Q_{2}^{*}:=Q_{2}^{*}(0) & =0,\nonumber 
\end{align}
$\left\langle \circ\right\rangle _{Q_{\alpha}^{\ast}}:=\left\langle \circ\right\rangle _{Q_{\alpha}^{\ast}(0)}(0)$,
$R=\sqrt{1+\gamma}\,\lambda_{\mathrm{max}}\approx\lambda_{\mathrm{max}}$
and $\gamma=\mathcal{O}(1/N)$ resulting from subleading terms in
\prettyref{eq:2ptcorreletor-1} containing $\mu$. \prettyref{eq:2ptcorreletor-1}
shows that $Q_{1}^{*}$ leads to a renormalization of the noise $D$
as $D_{\lambda}=D+Q_{1}^{*}=D/(1-R^{2})\approx D/(1-\lambda_{\mathrm{max}}^{2})$.

Analogously, second derivatives of saddle-points with respect to sources
can be calculated from \prettyref{eq:self-consis} with the final
result
\begin{eqnarray*}
\left.\frac{d^{2}Q_{1}^{\ast}(J)}{dJ_{k}dJ_{l}}\right|_{J=0} & = & \frac{\lambda_{\mathrm{max}}^{2}}{\beta N}\sum_{i}\left\langle X_{i}X_{k}\right\rangle _{Q_{\alpha}^{\ast}}\left\langle X_{i}X_{l}\right\rangle _{Q_{\alpha}^{\ast}}+\frac{\lambda_{\mathrm{max}}^{2}}{\beta N}\sum_{i,a}\left\langle X_{i}X_{a}\right\rangle _{Q_{\alpha}^{\ast}}^{2}\left.\frac{d^{2}Q_{2}^{\ast}(J)}{dJ_{k}dJ_{l}}\right|_{J=0},\\
\left.\frac{d^{2}Q_{2}^{\ast}(J)}{dJ_{k}dJ_{l}}\right|_{J=0} & = & \frac{\lambda_{\mathrm{max}}^{2}}{\beta N}\sum_{i}\left\langle \tilde{X}_{i}X_{k}\right\rangle _{Q_{\alpha}^{\ast}}\left\langle \tilde{X}_{i}X_{l}\right\rangle _{Q_{\alpha}^{\ast}},\qquad\text{with}\:\beta=1-\frac{\lambda_{\mathrm{max}}^{2}}{N}\sum_{i,a}\left\langle X_{i}\tilde{X}_{a}\right\rangle _{Q_{\alpha}^{\ast}}^{2},
\end{eqnarray*}
where we used that $\left.\frac{dQ_{\alpha}^{\ast}(J)}{dJ_{k}}\right|_{J=0}=0$
since three-point correlators vanish.

\selectlanguage{australian}%
Using \foreignlanguage{american}{partial derivative calculus $\frac{d}{dJ_{i}}\left\langle Z(J)\right\rangle =\left(\frac{\partial Q_{1}^{*}}{\partial J_{i}}\frac{\partial}{\partial Q_{1}^{*}}+\frac{\partial Q_{2}^{*}}{\partial J_{i}}\frac{\partial}{\partial Q_{2}^{*}}+\frac{\partial}{\partial J_{i}}\right)\left\langle Z(Q_{1}^{*},Q_{2}^{*},J)\right\rangle $
and the saddle-point condition $\frac{\partial}{\partial Q_{\alpha}^{*}}\left\langle Z(Q_{1}^{*},Q_{2}^{*},J)\right\rangle =0$
which follows from \prettyref{eq:saddle_1-1-1} and \prettyref{eq:Z-Y},
we find that second moments of activity, the average covariances,
are not influenced by the source dependence of saddle points
\begin{equation}
\left\langle \left\langle X_{i}X_{j}\right\rangle _{x}\right\rangle =\left.\frac{d^{2}}{dJ_{i}dJ_{j}}\left\langle Z(J)\right\rangle \right|_{J=0}=\left\langle X_{i}X_{j}\right\rangle _{Q_{\alpha}^{\ast}}=\left[\left(1-\mu\right)^{-1}D_{\lambda}\left(1-\mu^{\mathrm{T}}\right)^{-1}\right]_{ij}.\label{eq:second_moment_general}
\end{equation}
The non-zero mean connection strength $\mu=\mathcal{O}(1/\sqrt{N})$
yields cross-covariances between neurons due to the finite size of
the network. In addition, the fourth moments of activity are crucial
for the non-vanishing variance of covariances: Formally, two of the
four derivatives act on $Q^{*}(J)$ while the other two act on the
source term $J^{\mathrm{T}}X$ to yield
\begin{eqnarray}
\left\langle \langle X_{i}X_{j}X_{i}X_{j}\rangle_{x}\right\rangle  & = & \left.\frac{d^{4}}{dJ_{i}dJ_{j}dJ_{i}dJ_{j}}\left\langle Z(J)\right\rangle \right|_{J=0}\nonumber \\
 & = & 4\left.\frac{d^{2}Q_{1}^{*}(J)}{dJ_{i}dJ_{j}}\right|_{J=0}\sum_{a}\left\langle X_{i}\tilde{X}_{a}\right\rangle _{Q_{\alpha}^{\ast}}\left\langle X_{j}\tilde{X}_{a}\right\rangle _{Q_{\alpha}^{\ast}}+2\left.\frac{d^{2}Q_{1}^{*}(J)}{dJ_{i}dJ_{i}}\right|_{J=0}\sum_{a}\left\langle X_{j}\tilde{X}_{a}\right\rangle _{Q_{\alpha}^{\ast}}\left\langle X_{j}\tilde{X}_{a}\right\rangle _{Q_{\alpha}^{\ast}}\nonumber \\
 & + & 4\left.\frac{d^{2}Q_{2}^{*}(J)}{dJ_{i}dJ_{j}}\right|_{J=0}\sum_{a}\left\langle X_{i}X_{a}\right\rangle _{Q_{\alpha}^{\ast}}\left\langle X_{j}X_{a}\right\rangle _{Q_{\alpha}^{\ast}}+2\left.\frac{d^{2}Q_{2}^{*}(J)}{dJ_{i}dJ_{i}}\right|_{J=0}\sum_{a}\left\langle X_{j}X_{a}\right\rangle _{Q_{\alpha}^{\ast}}\left\langle X_{j}X_{a}\right\rangle _{Q_{\alpha}^{\ast}}\nonumber \\
 & + & 2\left\langle X_{i}X_{j}\right\rangle _{Q_{\alpha}^{\ast}}\left\langle X_{i}X_{j}\right\rangle _{Q_{\alpha}^{\ast}}+\left\langle X_{i}X_{i}\right\rangle _{Q_{\alpha}^{\ast}}\left\langle X_{j}X_{j}\right\rangle _{Q_{\alpha}^{\ast}}.\label{eq:fourth_moment_general-1}
\end{eqnarray}
The latter two terms in \prettyref{eq:fourth_moment_general-1} correspond
to the trivial Wick decomposition from the Gaussian part of the theory
at $Q_{\alpha}^{*}=Q_{\alpha}^{*}(J=0)$, whereas the second derivatives
of saddle points $\left.\frac{d^{2}Q_{\alpha}^{*}(J)}{dJ_{i}dJ_{j}}\right|_{J=0}=\mathcal{O}(1/N)$
evaluated at zero source determine the non-vanishing fourth cumulants.
Note that no quartic derivatives of saddle points appear in \prettyref{eq:fourth_moment_general-1}
due to the saddle-point condition $\frac{\partial}{\partial Q_{\alpha}^{*}}\left\langle Z(Q_{1}^{*},Q_{2}^{*},J)\right\rangle =0$.}
\selectlanguage{american}%

\subsection{Mean and variance of the covariance distribution}

We obtain the mean integral covariances (see \prettyref{eq:second_moment_general})
\begin{eqnarray}
\overline{c_{ij}} & = & \left[\left(1-\mu\right)^{-1}D_{\lambda}\left(1-\mu^{\mathrm{T}}\right)^{-1}\right]_{ij}=D_{\lambda}\gamma_{ij}\label{eq:mean_cij}
\end{eqnarray}
and the variance of integral covariances (see \prettyref{eq:c_ii2_bar-1},
\prettyref{eq:c_ij2_bar-1} and \prettyref{eq:fourth_moment_general-1})
\begin{eqnarray}
\overline{\delta c_{ij}^{2}} & = & \lambda_{\mathrm{max}}^{2}\left[\frac{1}{\left(1-\lambda_{\mathrm{max}}^{2}\right)^{2}}+\frac{1}{1-\lambda_{\mathrm{max}}^{2}}\right]D_{\mbox{\ensuremath{\lambda}}}^{2}\chi_{ij}.\label{eq:var_cij}
\end{eqnarray}
The mean connectivity $\mu$ enters the coefficients $\gamma_{ij}=\delta_{ij}+\gamma$
(with $\delta_{ij}$ the Kronecker symbol), $\chi_{ij}=\frac{1}{N}\left(1+\delta_{ij}+\mathcal{O}(1/N)\right)$
and $\gamma=\mathcal{O}(1/N)$ only in their sub-leading corrections,
and acts as a negative feedback in inhibitory or inhibition-dominated
networks \citep{Tetzlaff12_e1002596}. While this feedback suppresses
mean cross-covariances \eqref{eq:mean_cij} which consequently scale
as $\overline{c_{ij}}\sim\frac{1}{N}$, it only yields a subleading
contribution to the dispersion \eqref{eq:var_cij}. The spread of
individual cross-covariances is determined by fluctuations in connection
weights and shows a scaling as $\sqrt{\overline{\delta c_{ij}^{2}}}\sim\frac{1}{\sqrt{N}}$.
These fluctuations, which formally originate from the variability
of the auxiliary fields $Q$, are therefore much larger than the mean;
they cause broad distributions of cross-covariances of both signs
even in a homogeneous network. The expressions therefore explain the
first two moments of the experimentally observed distribution of cross-covariances:
mean cross-covariances scale as $\mathcal{O}(1/N)$, whereas the standard
deviation only scales as $\mathcal{O}(1/\sqrt{N})$. The width of
the distribution is thus much larger than the mean for large networks
and the distribution is centered approximately around zero. 

We note that this approach is inherently different from mean-field
theories for single realizations of network connectivities which keep
the site dependence to infer relations between covariances and connections
on the level of individual neurons \citep{Roudi11}. In contrast,
we derive a relation between the statistics of the structure and the
statistics of the dynamics using a quenched average. The crucial feature
of the presented theory is that heterogeneity across neurons can still
be extracted from the ensemble description. We here show that the
heterogeneity is closely linked to fluctuations of the auxiliary fields,
which are accessible by studying their source dependence.

\section{Bias and error of estimating the dispersion in experimental data}

\begin{figure}
\begin{centering}
\includegraphics{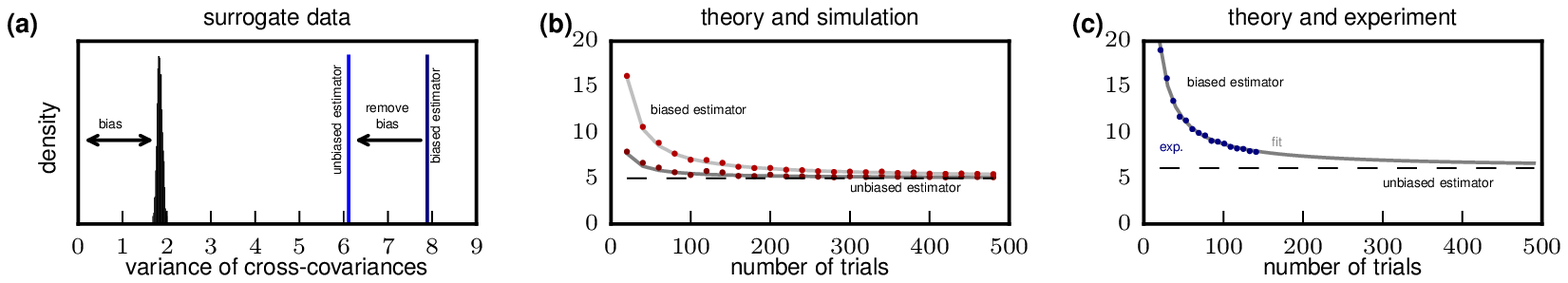}
\par\end{centering}
\caption{\textbf{Assessment and correction of the bias in estimator of variance
of cross-covariances.} (a) Variance of cross-covariances (black distribution)
obtained from trial-shuffled data compared to the biased estimator
(\prettyref{eq:biased_estimator}, dark blue) and unbiased estimator
(light blue) of the true variance as obtained from experimental data.
(b) Convergence of biased estimator for small (dark gray) and large
(light gray) mean covariances towards true variance given by unbiased
estimator (dashed line). Curves indicate the theoretical prediction
\prettyref{eq:biased_estimator}, dots numerical simulations for artificial
data. (c) True value $6.11$ of variance of cross-covariances (dashed
line) derived from fitting (gray curve) \prettyref{eq:biased_estimator}
to the biased estimator (blue dots) obtained from subsampled experimental
data with different numbers of trials. \label{fig:finite_data}}
\end{figure}
Due to limited numbers of trials, the estimator for the dispersion
of the covariance may be biased towards larger variances. We seek
to find a correction for this bias, as illustrated in \prettyref{fig:finite_data}a.
To this end we derive the dependence of experimentally estimated moments
of the activities on the number of neurons and trials. The derivation
here follows the standard approach of correcting the estimation, sometimes
called Bessel's correction \citep{Barlow13}.

We assume a probability distribution $p(n^{1},...,n^{N_{T}})$, $n^{k}\in\mathbb{N}_{0}^{N}$,
of activities $n_{i}^{k}$ of neuron $i$ in trial $k\in\{1,...,N_{T}\}$.
We assume that the spike counts are sufficiently large so that their
statistics is to leading order described by its first two cumulants,
which is in line with our findings in \prettyref{sec:Mean-field-theory-for}.
We further assume that there are no correlations between different
trials, so that the probability distribution factorizes over trials
and the moment generating function reads
\begin{equation}
\phi(l^{1},...,l^{N_{T}})=\prod_{k=1}^{N_{T}}\exp\left(m^{\mathrm{T}}l^{k}+\frac{1}{2}l^{k,\mathrm{T}}cl^{k}\right),\label{eq:moment_gen_func}
\end{equation}
where $m$ is the vector of mean activities with mean $\overline{m}$
and variance $\overline{\delta m^{2}}$ across neurons, $c$ is the
covariance matrix with mean autocovariance $\overline{a}$, variance
of autocovariances $\overline{\delta a^{2}}$, mean cross-covariance
$\overline{c}$ and variance of cross-covariances $\overline{\delta c^{2}}$
across neurons. The meta-statistics are assumed to be identical across
trials. Furthermore, we assume the meta-statistics to be the same
for all realizations of distributions of mean activities and covariances
across neurons. In the following, $\left\langle \right\rangle $ denotes
the average over these realizations obtained from the averaged moment
generating function 
\begin{eqnarray}
\left\langle \phi(l^{1},...,l^{N_{T}})\right\rangle  & = & \exp\left(\sum_{k=1}^{N_{T}}\left(\overline{m}\sum_{i}l_{i}^{k}+\frac{1}{2}\overline{a}\sum_{i}l_{i}^{k}l_{i}^{k}+\frac{1}{2}\overline{c}\sum_{i\neq j}l_{i}^{k}l_{j}^{k}\right)\right.\label{eq:average_moment}\\
 &  & \phantom{\exp}\left.+\sum_{k,l=1}^{N_{T}}\left(\frac{1}{2}\overline{\delta m^{2}}\sum_{i}l_{i}^{k}l_{i}^{l}+\frac{1}{8}\overline{\delta a^{2}}\sum_{i}l_{i}^{k}l_{i}^{k}l_{i}^{l}l_{i}^{l}+\frac{1}{8}\overline{\delta c^{2}}\sum_{i\neq j}l_{i}^{k}l_{j}^{k}l_{i}^{l}l_{j}^{l}\right)\right)\nonumber 
\end{eqnarray}
and $\hat{}$ denotes the empirical estimates of mean activities and
covariances from $N$ recorded neurons in $N_{T}$ trials of the experiment.
Note that the average across realizations formally introduces correlations
between different trials (mixed $k,l$ terms in \eqref{eq:average_moment})
which allow us to calculate corrections due to the finite number of
trials. Using \prettyref{eq:average_moment}, it is straight-forward
to show and well known that an empirical covariance defined as $\hat{c}_{ij}^{\mathrm{b}}=\frac{1}{N_{T}}\sum_{k=1}^{N_{T}}(n_{i}^{k}-\hat{m}_{i})(n_{j}^{k}-\hat{m}_{j})$
yields a biased estimator:
\begin{eqnarray}
\left\langle \hat{c}_{ij}^{\mathrm{b}}\right\rangle  & = & \frac{1}{N_{T}}\sum_{k=1}^{N_{T}}\left\langle (n_{i}^{k}-\hat{m}_{i})(n_{j}^{k}-\hat{m}_{j})\right\rangle =\frac{1}{N_{T}}\sum_{k=1}^{N_{T}}\left\langle n_{i}^{k}n_{j}^{k}\right\rangle -\left\langle \hat{m}_{i}\hat{m}_{j}\right\rangle =\frac{1}{N_{T}}\sum_{k=1}^{N_{T}}\left\langle n_{i}^{k}n_{j}^{k}\right\rangle -\frac{1}{N_{T}^{2}}\sum_{k,l=1}^{N_{T}}\left\langle n_{i}^{k}n_{i}^{l}\right\rangle \nonumber \\
 & = & \frac{N_{T}-1}{N_{T}}\left(\delta_{ij}\overline{a}+(1-\delta_{ij})\bar{c}\right)\label{eq:bias_mean}
\end{eqnarray}
with $\hat{m}_{i}=\frac{1}{N_{T}}\sum_{k=1}^{N_{T}}n_{i}^{k}$ and
$\left\langle n_{i}^{k}n_{j}^{l}\right\rangle =\delta_{ij}(\delta_{kl}\overline{a}+\overline{\delta m^{2}})+(1-\delta_{ij})\delta_{kl}\overline{c}+\overline{m}^{2}$.
The unbiased estimator is therefore given by $\hat{c}_{ij}=\frac{N_{T}}{N_{T}-1}\hat{c}_{ij}^{\mathrm{b}}=\frac{1}{N_{T}-1}\sum_{k=1}^{N_{T}}(n_{i}^{k}-\hat{m}_{i})(n_{j}^{k}-\hat{m}_{j})$
\citep{Barlow13}. Using this definition, along the same lines a
lengthy, but straightforward analogous calculation shows that the
variance of cross-covariances ($i\neq j$) defined as $\hat{\overline{\delta c^{2}}}=\frac{1}{N(N-1)}\sum_{i\neq j=1}^{N}(\hat{c}_{ij}-\hat{\overline{c}})^{2}$
with the mean cross-covariance across neurons $\hat{\overline{c}}=\frac{1}{N(N-1)}\sum_{i\neq j=1}^{N}\hat{c}_{ij}$
yields the biased estimator 
\begin{eqnarray}
\left\langle \hat{\overline{\delta c^{2}}}\right\rangle  & = & \left(1-\frac{2}{N(N-1)}\right)\left(\overline{\delta c^{2}}+\frac{\overline{a}^{2}-\overline{c}^{2}}{N_{T}-1}\right)\label{eq:biased_estimator}
\end{eqnarray}
of the true variance of cross-covariances $\overline{\delta c^{2}}$.
For a finite number of trials, there is a significant bias of $\left\langle \hat{\overline{\delta c^{2}}}\right\rangle $
caused primarily by the average variance $\overline{a}$ of spike-counts
across trials (see \prettyref{fig:finite_data}a for an empirical
estimate of the bias with trial-shuffled data).

The aim is hence to correct for the bias of the estimator. This can
be formally done using \prettyref{eq:biased_estimator}, which we
validate for synthetic data in\prettyref{fig:finite_data}b. A fit
of \prettyref{eq:biased_estimator} to the variance of cross-covariances
obtained from subsampled numbers of trials of the experimental data
shows that the true variance of cross-covariances can be approximately
obtained by subtracting the mean of the distribution of the surrogate
data in \prettyref{fig:finite_data}a.

In conclusion, the finite number of trials yields a significant bias
to the variance of cross-covariances, but not to the mean autocovariances,
if properly defined (see \prettyref{eq:bias_mean}). Fitting data
to \prettyref{eq:biased_estimator} and extrapolating for $N_{T}\to\infty$
yields the unbiased estimate which we use in the main text. However,
even neglecting this correction, the order of magnitude of the ratio
between the two estimates, which determines the largest eigenvalue
(Eq. (4) in the main text) and the operational regime, is not changed
by the bias.

In addition to the bias in the estimation of the variance, the estimator
comes with a statistical uncertainty due to the limited number of
recorded neurons. We observe $n=N(N-1)$ cross-covariances of which
we want to estimate the true variance. The relative standard error
of the variance is therefore given by $\sqrt{2/(n-1)}\approx0.009<1\%$
\citep{Lehmann06}. Error propagation to the estimation of $R^{2}$
shows that the relative error of $R^{2}$ is even smaller for $R^{2}>1/3$,
and hence in the critical regime close to $R^{2}\simeq1$ this error
is negligible.

Qualitatively similar distributions of correlations have been obtained
for other cortical areas, e.g. in visual cortex of macaque \citep{Ecker10}.
In this study, the authors consider correlations coefficients $\hat{z}_{ij}$
rather than covariances $\hat{c}_{ij}$, and showed that mean correlations
are close to zero. Furthermore, they showed a substantial contribution
to the dispersion of correlation coefficients arising from finite
data (see Fig. S3 in \citep{Ecker10}). The remaining variance after
this bias correction can be compared to our data if we re-interpret
the spike counts $n_{i}^{k}$ in the above calculations by the normalized
spike counts $n_{i}^{k}\rightarrow\frac{n_{i}^{k}}{\sqrt{\hat{c}_{ii}}}$.
Then the same derivation as above holds and we obtain an expression
for the width of the biased estimator of the variance $\hat{\overline{\delta z^{2}}}$
of correlation coefficients in terms of the true variance $\overline{\delta z^{2}}$
of correlation coefficients 
\begin{eqnarray}
\left\langle \hat{\overline{\delta z^{2}}}\right\rangle  & = & \left(1-\frac{2}{N(N-1)}\right)\left(\overline{\delta z^{2}}+\frac{1-\overline{z}^{2}}{N_{T}-1}\right).
\end{eqnarray}
For macaque motor cortex, we obtain a mean correlation coefficient
$\overline{z}=0.007$ and a standard deviation $\delta z_{ij}=\sqrt{\overline{\delta z^{2}}}=0.10$.
Both values are on the same order of magnitude as in \citep{Ecker10}
($\overline{z}=0.01$, $\delta z_{ij}=0.06$), but the dispersion
is larger in motor cortex. This motivates more detailed future investigations
of the distributions of correlations in various areas in relation
to their effective network size.

\section{Numerical simulations}

\subsection{Figure 2}

To illustrate the general mechanism relating network heterogeneity,
eigenvalues, and distributions of covariances, we consider a simple
network model with a sparsely and randomly connected inhibitory population
of size $N=1000$ and covariance of single-unit fluctuations $D=1$.
Connections $W_{ij}\stackrel{\mathrm{i.i.d.}}{\sim}\mathcal{B}(p,w)$
are drawn independently from the same Bernoulli distribution with
connection probability $p=0.1$ and weight $w=-1/\sqrt{N}$ (first
row), $w=-2/\sqrt{N}$ (second row), $w=-2.5/\sqrt{N}$ (third row)
and $w=-3/\sqrt{N}$ (fourth row).

\subsection{Figure 3}

All networks have sparse random connections of uniform strength drawn
from Bernoulli distributions. Furthermore, neurons have a fixed number
of incoming connections (in-degree) and do neither connect to themselves
(no autapses) nor are they connected multiple times to other neurons
(no multapses). The maximum eigenvalue on the abscissa (panels A-C)
is varied by the choice of the connection strength.

\paragraph{Homogeneous inhibitory network:}

We consider inhibitory populations of size $N=1000$ (panels A,B)
or $N=10000$ (panels C,E), $D=1$, connection probability $p=0.1$
and uniform non-zero weights varied in the range $w=-0.1,...,-0.001$
(panels A,B) or $w=-0.03,...,-0.003$ (panel C). Panel E shows data
for $w=-0.0285$.

\paragraph{Excitatory-inhibitory network:}

We consider a network of $N_{E}=8000$ excitatory and $N_{I}=2000$
inhibitory neurons with $D=1$, connection probability $p=0.1$, uniform
excitatory connection strengths varied in the range $w_{E}=0.001,...,0.01$
and uniform inhibitory connection strengths varied in the range $w_{I}=-0.06,...,-0.006$.
Panel F shows data for $w_{E}=0.009$ and $w_{I}=-0.05$. 

\paragraph{Inhibitory network with distance-dependent connectivity:}

We consider a network of $N=10000$ inhibitory neurons ($D=1$) randomly
positioned on a $1\mm\times1\mm$ sheet. Each neuron receives $K=100$
incoming connections of uniform strength varied in the range $w=-0.1,...,-0.01$.
Connections are drawn from a connection profile $p(x)\sim\exp(-x^{2}/(2\sigma_{\mathrm{conn}}^{2}))$
where $x$ is the Euclidean distance between the presynaptic and postsynaptic
neuron, and $\sigma_{\mathrm{conn}}=50\mum$ the space constant. 

\paragraph{Leaky integrate-and-fire network:}

We consider a network of $N=10000$ inhibitory leaky integrate-and-fire
model neurons with delta-shaped postsynaptic currents. The membrane
potential of each neuron follows the differential equation
\begin{eqnarray*}
\tau_{m}\frac{dV_{i}}{dt} & = & -V_{i}+\tau_{m}\sum_{j}J_{ij}s_{j}\left(t-h\right),
\end{eqnarray*}
where $s_{j}\left(t\right)=\sum_{k}\delta\left(t-t_{k}^{j}\right)$
is the spike train of the $j$-th neuron and $t_{k}^{j}$ denotes
the $k$-th spike of neuron $j$, which occurs whenever $V_{j}$ exceeds
the threshold $\theta$. The membrane potential is reset $V_{j}(t_{k}^{j}+)\leftarrow V_{r}$
to the reset potential $V_{r}$ after each such event and held at
this level for the absolute refractory time $\tau_{r}$. The time
lag $h$ equals the resolution of the time-driven simulation. We choose
a connection probability $p=0.1$ and uniform weights $J\in[-1.1,-0.1]$.
Neuron and simulation parameters are shown in \prettyref{tab:Parameters}.

\begin{table}
\begin{centering}
\begin{tabular}{@{\hspace*{1mm}}p{2.7cm}@{}@{\hspace*{1mm}}p{3.65cm}@{}@{\hspace*{1mm}}p{6.6cm}}
\hline 
\multicolumn{3}{>{\columncolor{parametergray}}c}{\textbf{Neuron model}}\tabularnewline
\textbf{Name} & \textbf{Value } & \textbf{Description}\tabularnewline
\hline 
$\taum$ & $20\ms$ & membrane time constant\tabularnewline
$\taur$ & $2\ms$ & absolute refractory period\tabularnewline
$V_{\mathrm{r}}$ & $0\mV$ & reset potential\tabularnewline
$\theta$ & $15\mV$ & fixed firing threshold\tabularnewline
$\EL$ & $0\mV$ & leak potential\tabularnewline
\end{tabular}\\
\begin{tabular}{@{\hspace*{1mm}}p{2.7cm}@{}@{\hspace*{1mm}}p{3.65cm}@{}@{\hspace*{1mm}}p{6.6cm}}
\hline 
\multicolumn{3}{>{\columncolor{parametergray}}c}{\textbf{Simulation parameters}}\tabularnewline
\textbf{Name} & \textbf{Value } & \textbf{Description}\tabularnewline
\hline 
$h$ & $0.1\ms$ & simulation time step\tabularnewline
$T$ & $1000$s & simulation time after initial transients\tabularnewline
$T_{\mathrm{trial}}$ & $1$s  & time window to compute spike counts\tabularnewline
\end{tabular}
\par\end{centering}
\centering{}\caption{Specification of neuron and simulation parameters for network of LIF
model neurons shown in Fig. 3 of the main text.\label{tab:Parameters}}
\end{table}

\subsection{Figure 4}

Linear instability is determined by an eigenvalue close to the critical
line $\mathrm{Re}(\lambda)=1$. For the dynamically balanced network
close to the critical point (Fig. 4 left) we consider a sparse, random
network of $N=1000$ inhibitory neurons with independent and identically
distributed connections $W_{ij}\stackrel{\mathrm{i.i.d.}}{\sim}\mathcal{B}(p,w)$.
This network by definition has an inhibition-dominated feedback. The
connection probability $p=0.1$ and weight $w=-3.1/\sqrt{N}$ are
chosen such that the theoretical prediction for the largest bulk eigenvalue
is $\lambda_{\mathrm{max}}=\sqrt{Np(1-p)w^{2}}=0.93\lessapprox1$.
Due to the finite size of the network, the largest real eigenvalue,
which is used in Fig. 4C, slightly differs from this result ($\lambda\approx0.917$).

For networks with almost vanishing feedback (Fig. 4 right), we consider
a sparse, random network of $N=1000$ excitatory neurons with a feedback
of order $1\ll N$, where $N$ is the network size. Connections $W_{ij}\stackrel{\mathrm{i.i.d.}}{\sim}\mathcal{B}(p,w)$
are independent and identically distributed. The connection probability
$p=0.1$ and weight $w=9.17/N$ are chosen such that the feedback
$N\mu=Npw=0.917\lessapprox1$ almost compensates the neuronal leak
(\prettyref{eq:OUP_functional}). The critical eigenvalue in this
network is given by the feedback $N\mu$ and corresponds to the population
activity.
\end{document}